\documentstyle[12pt]{article}
\begin{document}
\hspace*{11.2cm}HU-TFT-96-13
\vskip 4.5 cm
\begin{center}
{\bf Dangers of Unphysical Regions}
\end{center}
\vskip 1.0 cm
\begin{center}
\renewcommand{\thefootnote}{a)}
Matts Roos\footnote{Laboratory of High Energy Physics, P.O.B. 9,
FIN--00014 UNIVERSITY OF HELSINKI, Finland.}
and \renewcommand{\thefootnote}{b)}Leonid A. Khalfin 
\footnote{Russian Academy of Sciences, Steklov Mathematical 
Institute, St. Petersburg Department, Fontanka 27, St. Petersburg 191011,
Russia.} \\

\end{center}

\begin{abstract} 
We discuss the appearance of negative numbers of events in radiochemical
experiments and negative squared antineutrino mass $m^2_{\bar{\nu}_e}$ in
tritium beta decay. Going beyond the standard discussion about how to extract
upper limits in those cases, we show that the problem is much more profound.
We explain the circumstances which are the likely cause of the persistently
negative values of $m^2_{\bar{\nu}_e}$ in all modern tritium beta decay
experiments.      

\end{abstract}
\newpage

{\bf Introduction.}

A problem in data analysis which turns up occasionally in some experiments and
notoriously in others is that of a parameter estimate occurring outside the
theoreti\-cal\-ly allowed physical range of the parameter. 
The common attitude to this is that one may use the unphysical result to set a
limit inside the physical range. If one assumes that a known, $e.g.$ Gaussian, 
probability density function can be centered on the measured value found in the
unphysical range, there exists a standard prescription for translating a
confidence interval for the measured value into a confidence interval for
the true value \cite{Eadie,knapp}. The discussion then generally turns around
whether one should use the full Gaussian for the confidence interval, or only
the part of the Gaussian tail which is inside the physical range, properly
renormalized \cite{knapp,james,pdg}. 

For this discussion we shall use two well known examples where the problem with
unphysical regions appear. The first example is the distinction of a positive
signal over background, where both signal and background consist of countable
events. Our second example is the case of neutrino mass determination from the
shape of the electron energy spectrum in tritium $\beta$-decay. The neutrino
mass example was also used in refs. \cite{knapp,james}, where it was assumed
that a Gaussian could meaningfully be centered on the estimated value in the
unphysical range. In this paper we shall argue that the problem is more
profound, and that the data analysis should be done differently.

\bigskip
{\bf Radiochemical experiments.}

Consider the detection of radiatively decaying atoms in a radiochemical 
experiment, such as the solar neutrino experiments
\cite{Homestead,Gallex,Sage}. The radioactive atoms have been produced by
exposure to the neutrino radiation from the sun, and the total number of atoms
collected gives information about the solar neutrino flux. The atoms are
counted when the radioactive decay takes place in a detector with well known
sensitivity, and the signature of these events is that they occur with a well
known time constant $\lambda$. There are no other characteristics permitting
the distinction of signal events from background events occurring randomly in
time. The estimator of the sum of neutrino flux and background is thus a
Poisson-distributed discrete random  variable. It is essential to note that the
number of counts is very small during a period of observation, a "run",
sometimes zero. The general procedure for how to analyze such data has been
well described by Cleveland \cite{Cleveland}.
        
The signal events are defined to occur with a time distribution $\exp (-\lambda
t)$, and the background is assumed constant in time. Thus the occurence
probability per unit time is given by
\begin{eqnarray}
f(t;a,b)=ae^{-\lambda t} + b\ ,
\label{f1}\end{eqnarray}
where $a$ and $b$ are the signal and background intensities, respectively. 
Since both the signal and the background are Poisson-distributed non-negative
numbers the physical region is defined by $a$ and $b$ both being non-negative.

The estimators $\widehat{a}$ and $\widehat{b}$ are found by maximizing the
likelihood function (or minimizing the negative of the log-likelihood function)
for $N$ observations of one event, each at a time $t_i, \ i=1...N$,
\begin{eqnarray}
{\cal{L}}(t_1,...,t_N\vert a,b)=e^{-a\Delta/\lambda -bT}
\prod_{i=1}^N (ae^{-\lambda t_i} + b)\ ,
\end{eqnarray}
where $T$ is the total observation time, and $\Delta$ is the sum of the $N$
time intervals weighted by $\lambda$,
\begin{eqnarray}
\Delta=\sum_{i=1}^N e^{-t_i\lambda}-e^{-t_{i-1}\lambda}\ .
\end{eqnarray}
Usually $\widehat{a}$ and $\widehat{b}$ are not required to be separately
non-negative.  If $\widehat{a}$ and $\widehat{b}$ are significantly
positive (as in the total fits to a large number of independent experimental
"runs") this causes no problem. 

Consider, however, the situation in an indvidual run when the number of signal
events is zero or nearly zero. If $\widehat{a}$ is not required to be
non-negative in the fit, the maximum likelihood may occur for a value
of $\widehat{a}$ which is negative as a result
of a pure statistical fluctuation. 

Or else, there may be an unknown process contributing to the background so that
the data happens to exhibit a component increasing in time, for instance a late
accumulation of events which simulates an intensity increasing with time. Since
this is not taken care of by the $b$ assumed constant, this is misinterpreted
as a signal having the form $f(t;-a,b)$. In other words, {\it the background
hypothesis is wrong}.                        

Since there is no physical theory in an unphysical region, $f(t;-a,b)$ is
arbitrary, and could often be replaced by some other arbitrary continuation.
Thus if the likelihood function has a deeper extremum for some negative $a$,
this fact cannot be used to make confidence statements about $\widehat{a}$ in
the physical region, because the choice of another  continuation might yield a
negative $\widehat{a}$ corresponding to a different extremum. In any case one
should bear in mind that the information obtained using $f(t;-a,b)$ is an
information on the background, not on $a$.

A further problem is that $f(t;-a,b)$ is not a normalizable probability
distribution. It therefore does not make sense to describe the data by a
Gaussian centered on $-\widehat{a}$, and to make inferences from its tail in
the physical region. 

The only inference one can make about a negative signal is that
the hypothesis (\ref{f1}) is wrong, notably that the background is not well
described by a constant.  To prove that the hypothesis (\ref{f1}) is wrong one
has to make a goodness-of-fit test in the physical region, for instance at 
$a=0$, not a test which makes use of the arbitrary and ill-defined likelihood
in the unphysical region. Unfortunately the widely used chisquare test is not
reliable for very small numbers of events.

If the extremum is on the edge of the allowed parameter range there may be a
problem with the convergence of numerical search algorithms such as those used
in MINUIT \cite{MINUIT}. Note that the hypothesis used in an unphysical
region may influence the value of $\widehat{a}$ in the nearby physical  region,
because the search algorithms make finite steps around the extremum.

The distribution of a {\it consistent} estimator converges in the (weak) limit
of infinite volume of the sample to the delta function for the true value. An
{\it unbiassed} estimator, repeatedly measured, returns values on both sides of
the true value while converging towards it. However, when the true value is
exactly zero (an academic case) and the estimator is a Poisson-distributed
random variable restricted to integer non-negative values, the measurement
process can only converge from the positive side, thus it appears biassed.  But
this is unavoidable, because mathematical statistics does not admit any
estimator for negative values of a Poisson variable. If instead one does use an
{\it ad hoc} estimator, not prescribed by the physical theory, which can take
values on both sides of zero, the gain in apparent unbiassedness is obtained
at the price of arbitrariness.

Let us return to the case when a large number of independent experimental runs
are done. The discussion has generally concerned the question whether
$\widehat{a}$  should be constrained to be non-negative in the individual
runs before averaging. The answer to this is clear: if the results are
non-negative {\it on average}, one should not constrain the individual
runs  because that would bias the final result towards higher values of
$\widehat{a}$. Although this is not wrong, it implies the loss of  real
experimental information. The advisable procedure is to fit all the runs
simultaneously with one common parameter $\widehat{a}$ and with (if necessary) 
different background parameters $b_j$ for each run $j$. When the ensuing
average $\widehat{a}$ is in the physical region, no error in procedure would
have occurred, because also the runs which individually yielded negative
signals would be contributing their share to the total likelihood at the common
best value of $\widehat{a}$ in the physical region. This is the procedure used
for instance by GALLEX \cite{Gallex}.   
                                      
\bigskip
{\bf The neutrino mass problem.}

Consider now the case of the electron antineutrino mass determination from the
shape of the electron energy spectrum in tritium $\beta$-decay. The spectrum
given by Fermi theory is                           
\begin{eqnarray}
f(E\vert A,m^2,E_0,b)= ApE\ F(E,E_0)\ (E_0-E)\ [(E_0-E)^2-m^2]^{1\over 2}+b\ .
\label{f2}\end{eqnarray}
$F(E,E_0)$ is called the Fermi function (see $e.g.$ ref. \cite{knapp}), p is
the electron momentum, and the root is taken to be real and positive.
Note that the $e$-antineutrino mass $m$ enters only in the form $m^2$. The
other parameters to be determined by the fit are the total decay energy $E_0$,
a positive normalization constant $A$, and the background $b$ assumed constant.
(Sometimes the background is described by an empirical function, $e.g.\ \
b+cE$ \cite{Mainz}, but this does not affect our general argumentation.)

The physical region is defined by conditions on $m^2$ and $E_0$. Clearly $m^2$
must be non-negative. From the factor $(E_0-E)$ we see that the theoretical
electron spectrum would be negative for energies $E>E_0$, and the square root
becomes imaginary for $E>E_0-m$. Oddly enough, the square root term in
Eq.~(\ref{f2}) is positive and real for all energies $E<E_0$ if $m^2$ were
allowed to be negative. This curious fact is the cause of problems encountered
in all fits, as has been pointed out before \cite{Khalfin1}, and as we shall
discuss below.

The theory of beta decay does prescribe the electron energy spectrum to have
the form (\ref{f2}), but the atomic physics or molecular physics of the tritium
compound might cause modifications. The Fermi function depends on
approximations which we shall not discuss here. The experimental resolution
function has to be folded into the final spectrum, and all possible
instrumental distortions might not be well understood.  In any case, the
assumption \cite{Troitsk,LLL} that the response (resolution) function of the
$\beta$-spectrometer is Gaussian can not be rigorously proved for the tail of
the response function.

Thus the form of the signal is not well known and the hypothesis (\ref{f2}) as
well as the assumption for the background may be wrong. An accumulation of
events at energies near or beyond $E_0$ have been reported \cite{Troitsk,LLL}
indicating problems with Eq.~(\ref{f2}). The experimental background is not due
to tritium $\beta$-decay, and thus its distribution does not respect the
kinematical limit $E<E_0-m$.  
                 
The estimators $\widehat{m^2}$ and $\widehat{E_0}$ are random variables which
depend on the data and the method, $e.g.$ in all recent experimental works 
(see \cite{pdg}  and references therein, and refs. \cite{Troitsk,LLL}) the
method is least squares minimization, the "chisquare" method \cite{Eadie}.
               
The comments we made on the unphysical region in the radiochemical case are
mostly the same here, but there are important differences due to the more
complicated mathematical form of the "signal". From the expression (\ref{f2}) 
we obtain for $E=E_0-m$ if $m\neq 0$
\begin{eqnarray}
df((E_0-m)\vert A,m^2,E_0,b)/dm^2=\infty\ .
\end{eqnarray}

Moreover, taking into account that the Fermi function $F(E,E_0)$ has no
singularity at $E=E_0-m$ we obtain, if $m\neq 0$, from the expression
(\ref{f2}) for $E=E_0-m$  and any $n=1,2,...$
\begin{eqnarray}
\vert d^nf((E_0-m)\vert A,m^2,E_0,b)/dE^{n}\vert =\infty\ .
\end{eqnarray}
Thus there is no analytic continuation of the theoretical "signal" (\ref{f2})
from the  physical region  into the unphysical region. It follows that it does
not make sense to describe the data by a Gaussian centered at $-\widehat{m^2}$,
and make inferences from its tail in the physical region.

And even if it did make sense (putting in an empirical analytic continuation
"by hand" \cite{Troitsk,LLL} and thereby changing the given problem of
estimating the physical nonnegative $m^2$), one cannot derive a confidence
limit on $m$ from $-\widehat{m^2}$ \cite{james}.

There are two possible reasons for obtaining a negative $\widehat{m^2}$ in
a numerical search for minimum chisquare. The first reason is trivial, and
not of statistical nature: there is a real accumulation of events near the
end point $E=E_0-m$ caused by instrumental problems or unknown physics or a
wrong background hypothesis. Since Eq.~(\ref{f2}) does not account for this, it
must be modified. 

The conclusion is then the same as in the radiochemical case. A goodness-of-fit
test of the hypothesis (\ref{f2}) should be made in the physical region, not at
$-\widehat{m^2}$. There is again the caveat about the chisquare method not
being a suitable test in a region with very low statistics.

The second reason is quite non-trivial \cite{Khalfin1}.  No tritium decay
electron can have an energy $E>E_0-m$, so data obtained in that range must be 
background. Yet it is impossible to restrict the fit of Eq.~(\ref{f2}) to the
physical range $E\leq E_0-m$ and replacing it in the $E>E_0-m$ range by a pure
background term, because $E_0$ is an unknown and itself a parameter to be
determined in the fit.  And even if we knew from theory the approximate value
of $E_0$ (as was assumed in \cite{LLL}), the problem is the same, because we
do not know the antineutrino mass.             

This problem was already understood by the writers of MINUIT, who  built an
explicit check into the program, prohibiting the user from supplying different
functions in different parameter ranges if the ranges depend on one of the
parameters to be estimated \cite{MINUIT}.

Therefore, it is almost unavoidable that the fit covers data in some part of
the unphysical range $E>E_0-m$, where the numerical search $must$ make steps
into the region of negative $\widehat{m^2}$ in order to keep the root in (\ref{f2})
real. And, as already mentioned in the radiochemical case, search algorithms 
like those used in MINUIT \cite{MINUIT} only converge if they can make finite 
steps around the extremum, thus stepping $E_0$ occasionally into the range
$E_0<E+m$. The convolution of the energy resolution function into the spectrum
may also be a cause for excursions into this unphysical range.

Of course, this situation is avoidable if one has sufficiently good estimates
for $E_0$ and $m$ to be able to restrict the fit to experimental data in the
range $E\ll (E_0-m)$ far from the unphysical region,  as were done in the past
(see \cite{langer} and references therein). Although one then obtains
non-negative estimates for $\widehat{m^2}$, the resulting statistical errors
(dispersion) are big since the $\beta$-spectrum for $E$ far from $E=E_0-m$
is very insensitive to the unknown antineutrino mass.          

We believe that the described circumstances are the main reason why all modern
experiments \cite{pdg,Troitsk,LLL}  have obtained results in the unphysical
range $\widehat m^2<0$. We are currently carrying out simulations to
demonstrate this quantitatively \cite{Khalfin2}, but it would still be 
preferrable if the experimental groups would confirm this on real data.

\bigskip
{\bf Conclusions.}

For both cases studied, the determination of the solar neutrino flux  by
radiochemical methods and the determination of the electron antineutrino mass
from the end point of the electron spectrum in tritium $\beta$-decay,  the
conclusions are the same. An analysis which yields a signal estimate which is
zero or negative obviously tells that there is no signal, only background. 
Thus the information obtained from the unphysical region with a negative signal
is not an information on the signal parameter, but on the background. 

In the radiochemical case the functional form of the signal in the unphysical
region is an increasing function which approaches a constant. If the 
background happens to have this form, for physical reasons or because of
statistical fluctuations, a negative signal is the apparent result. The same is
true in the electron antineutrino mass case  but more importantly, the squared
antineutrino mass is forced to be negative by attempts to fit  the electron
spectrum and the background in the energy region beyond the physical end point
of the spectrum.
                             
In both cases, a fit which is better in the unphysical region than in the
physical region is an accident, because the function which is correct in the
physical region is arbitrary in the unphysical region, and could often be
replaced by some other arbitrary continuation. Thus an extremum of the
likelihood function in the unphysical region cannot be used to make confidence
statements about the parameter in the physical region, because the choice of
another arbitrary function would yield a different result.
                          
In particular it is impossible to obtain the confidence statement about the
real antineutrino mass from a negative unphysical estimator, $\widehat{m^2}<0$
\cite{james}. All one can conclude (barring mistakes on the functional form of
the background) is that $m\neq 0$, because if it were zero, there would not be
any square root in Eq.~(\ref{f2}), and nothing would force the fit
into the region of negative $\widehat {m^2}$ \cite{Khalfin1}.              

To decide whether the function composed of signal plus background is an
adequate description of the data, one has to make a goodness-of-fit test in the
physical region. It is misleading to make a likelihood ratio test comparing the
likelihood in an unphysical region with the likelihood in the physical region,
because the former is meaningless and the functions are not the same.  It is
doubtful whether  the chisquare method is a suitable goodness-of-fit test near
the edge of a physical region, because  chisquare is not reliable for small
samples. The resolution of this problem is to find other statistical tests
which are adapted to the situation.
    
We have shown that there are circumstances which cause  a fit to the electron
spectrum in tritium $\beta$-decay to yield a negative $m^2$ value. Since these
circumstances are present in all modern analyses \cite{pdg,Troitsk,LLL}, we
believe we have found the reason for the anomalous results.

\bigskip
We thank Fred James, CERN, for useful comments and valuable criticism.

\end{document}